\documentclass[aps,prb,twocolumn,superscriptaddress]{revtex4-2}
\usepackage[dvips]{graphicx}
\usepackage{color}
\usepackage{bm}
\usepackage{amssymb}

\usepackage[utf8]{inputenc}
\usepackage{times}

\usepackage{amsmath}
\usepackage{amssymb}
\usepackage{graphicx}
\usepackage{dcolumn}
\usepackage{bm}
\usepackage{textcomp}
\usepackage{braket}
\usepackage{comment}
\usepackage{siunitx}
\usepackage{upgreek}
\usepackage{mathrsfs}
\usepackage{multirow}
\usepackage{physics}
\usepackage{natbib,hyperref}
\usepackage{booktabs}
\usepackage{wasysym}
\usepackage{comment}
\usepackage[caption=false]{subfig}

\def\HGSO{Ho$_2$GaSbO$_7$}
\def\HSSO{Ho$_2$ScSbO$_7$}
\def\HTO{Ho$_2$Ti$_2$O$_7$}

\begin{document}

\title{Influence of controlled disorder on the dipolar spin ice state of Ho-based pyrochlores}

\author{Adam A. Aczel} \altaffiliation{\href{mailto:aczelaa@ornl.gov}{aczelaa@ornl.gov}}
\affiliation{Neutron Scattering Division, Oak Ridge National Laboratory, Oak Ridge, TN 37831, USA}

\author{Brenden R. Ortiz} 
\affiliation{Materials Science and Technology Division, Oak Ridge National Laboratory, Oak Ridge, TN 37831, USA}

\author{Yi Luo}
\affiliation{Neutron Scattering Division, Oak Ridge National Laboratory, Oak Ridge, TN 37831, USA}

\author{Ganesh Pokharel}
\affiliation{Materials Department, University of California Santa Barbara, Santa Barbara, CA 93106, USA}
\altaffiliation{Current affiliation: Perry College of Mathematics, Computing, and Sciences, University of West Georgia, Carrollton, Georgia 30118, USA}

\author{Paul M. Sarte}
\affiliation{Materials Department, University of California Santa Barbara, Santa Barbara, CA 93106, USA}

\author{Clarina dela Cruz}
\affiliation{Neutron Scattering Division, Oak Ridge National Laboratory, Oak Ridge, TN 37831, USA}

\author{Jue Liu}
\affiliation{Neutron Scattering Division, Oak Ridge National Laboratory, Oak Ridge, TN 37831, USA}

\author{Gabriele Sala} 
\affiliation{Oak Ridge National Laboratory, Oak Ridge, TN 37831, USA}
\altaffiliation{Current affiliation: Materials and Life Science Division, J-PARC Center, Japan Atomic Energy Agency, Tokai, Ibaraki, 319-1195, Japan}

\author{Stephen D. Wilson}
\affiliation{Materials Department and California Nanosystems Institute, University of California Santa Barbara, Santa Barbara, CA 93106, USA}

\author{Benjamin A. Frandsen} 
\affiliation{Department of Physics and Astronomy, Brigham Young University, Provo, UT 84602, USA}

\author{Joseph A.~M. Paddison} 
\affiliation{Neutron Scattering Division, Oak Ridge National Laboratory, Oak Ridge, TN 37831, USA}

\date{\today}

\begin{abstract}
Pyrochlore magnets of the form $R_2B_2$O$_7$, in which rare-earth ions on the $R$-site form a three-dimensional network of corner-sharing tetrahedra, provide a canonical setting for geometrical frustration. Ho-based pyrochlores host a dipolar spin-ice ground state, characterized by Ising moments constrained by the ice rules and elementary excitations analogous to magnetic monopoles. Here we examine how controlled chemical disorder influences this state by introducing site mixing on the non-magnetic $B$-site in two compounds. Ho$_2$GaSbO$_7$ contains only Ga$^{3+}$/Sb$^{5+}$ charge disorder, whereas Ho$_2$ScSbO$_7$ exhibits both charge and substantial size disorder arising from the large ionic-radius mismatch between Sc$^{3+}$ and Sb$^{5+}$. Although both materials retain the pyrochlore structure, neutron scattering measurements reveal a reduced correlation length for the $R/B$-site cation ordering and enhanced local structural distortions in Ho$_2$ScSbO$_7$. Despite these structural differences, bulk thermodynamic measurements and magnetic diffuse scattering demonstrate that both systems exhibit the defining signatures of a dipolar spin-ice state. Low-energy inelastic neutron spectroscopy further uncovers broad magnetic excitations that develop within the dipolar spin-ice regime, a feature absent in pristine Ho pyrochlores and indicative of disorder-induced splitting of the non-Kramers ground-state doublet. Together, these results show that controlled disorder generates tunable transverse-field-driven quantum fluctuations in Ho-based pyrochlores, although the dipolar spin-ice state is remarkably robust to this disorder. 
\end{abstract}

\maketitle

\section{Introduction}

Rare-earth pyrochlore oxides of the form $R_2 B_2$O$_7$, where $R$ is a magnetic trivalent rare-earth ion and $B$ is a non-magnetic tetravalent cation, host a variety of exotic magnetic ground states and phenomena \cite{06_greedan, 10_gardner, 15_wiebe, 18_hallas, 19_rau, 19_zhou, 25_smith}. The $R$-ions form a network of corner-sharing tetrahedra, which is one of the canonical frustrated lattices that can prevent all pairwise-interactions from being satisfied simultaneously. Ho and Dy-based Ising pyrochlores, with the moments constrained to local $\langle111\rangle$ directions, are of special interest due to their penchant for realizing a dipolar spin ice (DSI) ground state \cite{99_ramirez, 01_bramwell, 01_bramwell_2}. The nearest-neighbor (NN) spin ice Hamiltonian \cite{98_bramwell, 04_melko} captures much of the important DSI physics and is given by
\begin{equation}\label{Eq:Ham1}
\mathcal{H_\mathrm{NNSI}} = \sum_{\langle i,j\rangle} J_{\parallel} \sigma^z_i \sigma^z_j,
\end{equation}
where $\sigma^z$ is a pseudospin operator and $J_{\parallel}>0$ is an antiferromagnetic Ising interaction between nearest-neighbor pseudospins. Note that the pseudospin operators are written in the local reference frame of the pyrochlore lattice \cite{11_ross, 19_rau_2}. The DSI model has an extensive degeneracy of spin configurations satisfying the `ice rule' of two spins pointing in and two pointing out of each tetrahedron. The elementary excitations of DSI are gapped emergent magnetic monopoles corresponding to spin flips that generate deviations from the ice rules \cite{08_castelnovo, 12_castelnovo}. DSI systems typically freeze into glassy spin-ice states as the temperature is decreased well below the energy scale of $J_{\parallel}$, since these spin flip processes are no longer thermally accessible. Ho$_2$Ti$_2$O$_7$ and Dy$_2$Ti$_2$O$_7$ are dipolar spin ice systems that have been characterized extensively \cite{99_ramirez, 00_matsuhira, 01_bramwell_2, 01_matsuhira, 01_synder, 02_fennell, 02_fukazawa, 04_fennell, 05_fennell, 09_clancy, 09_fennell, 11_giblin}. 

Adding other terms to the NNSI model can generate a quantum spin liquid ground state. One approach is to introduce a transverse exchange term $J_\perp$ to stabilize `quantum spin ice' (QSI) \cite{11_ross, 12_benton, 12_shannon, 14_gingras}. The modified Hamiltonian is given by 
\begin{equation}\label{Eq:Ham2}
\mathcal{H_\mathrm{QSI}} = \sum_{\langle i,j\rangle} J_{\parallel} \sigma^z_i \sigma^z_j - \sum_{\langle i,j \rangle} J_{\perp} (\sigma^+_i \sigma^-_j + \sigma^-_i \sigma^+_j).
\end{equation}
The $J_\perp$ term promotes quantum tunneling between different ice configurations and prevents the spin freezing that is realized in DSI. The QSI model is characterized by three different flavors of excitations: gapless photon modes, gapped visons, and magnetic monopoles \cite{14_gingras}. Candidate QSI systems with non-zero $J_\perp$ terms include Ce \cite{19_gaudet, 19_gao, 20_sibille, 22_smith, 22_bhardwaj, 24_zhao, 25_smith, 25_gao, 25_smith_2, 25_poree} and Pr-based \cite{08_zhou, 13_kimura, 17_sarte, 18_sibille, 24_ortiz} pyrochlore oxides, although it has been difficult to identify this phase unambiguously in the laboratory.

An alternative approach to stabilizing a QSI state starting from the NNSI Hamiltonian is to consider the special case of non-Kramers magnetic ions and add chemical disorder via non-magnetic doping or strain to the model \cite{17_savary, 18_benton, 19_pardini, 25_marinho}. In non-Kramers systems, only $\sigma^z$ carries a dipolar magnetic moment, whereas $\sigma^x$ and $\sigma^y$ correspond to higher-order multipolar degrees of freedom. Any disorder that lowers the local $D_{3d}$ symmetry of the rare-earth ion lifts the ground-state doublet degeneracy and produces two singlets with a continuous distribution of energy splittings. This single-ion effect can be captured by introducing a distribution of random transverse fields (RTF), leading to the effective Hamiltonian
\begin{equation}\label{Eq:Ham3}
\mathcal{H}_\mathrm{RTF} = \sum_{\langle i,j \rangle} J_{\parallel} \sigma^z_i \sigma^z_j - \sum_i h_i \sigma^x_i .
\end{equation}
Here the site-dependent transverse field $h_i$ couples to $\sigma^x_i$ and quantifies the deviation from ideal $D_{3d}$ symmetry induced by local structural disorder. The disorder-induced quantum spin ice (DQSI) state is stabilized when an antiferromagnetic $J_{\parallel} > \frac{5}{3} \bar{h}$, where $\bar{h}$ corresponds to the average transverse field strength, and the distribution width of the transverse fields $\delta h$ is sufficiently small to prevent the formation of a spin-glass-like phase instead \cite{18_benton, 19_pardini}. A paramagnetic state is realized when $J_{\parallel} < \frac{5}{3} \bar{h}$. It has been pointed out that the magnetism of both Pr and Ho-based pyrochlores with chemical disorder should be well described by this model \cite{18_benton}. 

The random-transverse field model has been employed in select Pr-based pyrochlores to explain their magnetic properties \cite{17_wen, 25_luo}, since disorder appears to be prelevant in these systems. In fact, Pr$_2$Zr$_2$O$_7$ was initially proposed as a DQSI candidate with the disorder arising from off-centering of the Pr$^{3+}$ ions \cite{14_koohpayeh} from the ideal pyrochlore positions \cite{17_wen}, but it has since been suggested that the transverse fields are sufficiently large to induce paramagnetic behavior in a large volume fraction of the sample instead \cite{18_benton, 25_hicken}. More recently, the source of the Pr$^{3+}$ off-centering in Pr$_2$Zr$_2$O$_7$ has been attributed to Pr$^{3+}$ stuffing on the Zr$^{4+}$ site \cite{25_hicken} and evidence for proximate DQSI behavior has been elucidated in single crystalline Pr$_2$Sn$_2$O$_7$ \cite{25_luo}. Despite this progress, the intrinsic magnetic ground states of the Pr-based pyrochlores are still a matter of debate, as sample-dependent low-temperature properties (e.g. AC susceptibility and heat capacity) are prevalent \cite{02_matsuhira, 09_matsuhira, 13_kimura, 16_anand, 16_petit, 16_sibille, 22_tang, 24_ortiz, 25_luo}, perhaps due to variable, uncontrolled amounts of disorder. The relative importance of transverse exchange and random transverse fields in these materials is also poorly understood, particularly since there is no theoretical work that takes both terms into account simultaneously.

Ho-based pyrochlores with controlled disorder are ideal experimental platforms for investigating the predictions of the random transverse-field model, since the magnetic ground states of the disorder-free systems are established and they have no transverse exchange term in their magnetic Hamiltonians. While mixed $B$-site Ho-based pyrochlores offer one straightforward approach to introducing this controlled disorder, surprisingly little characterization work has been done on these systems. There are a few reports on Ho$_{2+x}$Ti$_{2-x}$O$_7$ `stuffed spin ice' \cite{07_lau, 08_ehlers, 13_aldrus} and Ho$_2$InSbO$_7$ \cite{ortiz2022traversing}, but the former involves doping magnetic ions on the $B$-site and the latter is argued to crystallize in the defect-fluorite structure, so neither case can be used as a valid test-bed for the random transverse field theory. 

Here, we report comprehensive measurements on the mixed $B$-site pyrochlore systems Ho$_2$GaSbO$_7$ and Ho$_2$ScSbO$_7$. We selected these compounds to facilitate a systematic investigation of the effects of chemical disorder induced by variation in the charge and ionic radii of the mixed $B$-site cations: Ga$^{3+}$ and Sb$^{5+}$ have very similar radii (0.62~\AA~and 0.60~\AA) and would thus be expected to produce relatively weak local disorder, while Sc$^{3+}$ is much larger (0.745~\AA), leading to stronger local disorder. Detailed structural characterization using neutron scattering confirms this scenario and highlights key structural differences between the two compounds, yet the magnetic characterization reveals a robust DSI state in both. Low-energy excitations probed via neutron spectroscopy shed further light on the influence of the local disorder on the random transverse fields. Taken together, these results provide a detailed picture of the chemical disorder in two mixed $B$-site pyrochlores, elucidate the influence of such disorder on the magnetic ground state, and demonstrate the robustness of the DSI state in disordered Ho pyrochlores.

\section{Methods}

Polycrystalline samples of Ho$_2$Ti$_2$O$_7$, Ho$_2$GaSbO$_7$, and Ho$_2$ScSbO$_7$ were made using the mechanicochemical method proposed in our synthesis of the \textit{R}$_2$InSbO$_7$ pyrochlore oxides previously \cite{ortiz2022traversing}. Stoichiometric amounts of dried Ho$_2$O$_3$ (99.995\%, Alfa Aesar), Ga$_2$O$_3$ (99.999\%, Alfa Aesar), Sc$_2$O$_3$ (99.999\% Alfa Aesar), TiO$_2$ (99.99\% Alfa Aesar), and Sb$_2$O$_5$ (99.998\% Alfa Aesar) were combined into a tungsten carbide ball-mill vial and milled for 1~h. The resulting powders were extracted, ground in an agate mortar, and sieved through a 50~$\mu$m sieve. These powders were loaded into 2~mL, high-density alumina crucibles (CoorsTek) and annealed in a box furnace at 1250$^\circ$C for 48~h.

Time-of-flight neutron powder diffraction (NPD) data were collected at 115~K using the NOMAD beam line \cite{neuef;nimb12} at the Spallation Neutron Source (SNS) of Oak Ridge National Laboratory (ORNL) on Ho$_2$Ti$_2$O$_7$, Ho$_2$GaSbO$_7$, and Ho$_2$ScSbO$_7$. The powder samples (typical mass $\sim$ 0.3 g) were loaded into thin quartz capillaries and mounted on the beamline, with a nitrogen cryostream used to cool the samples. Standard data processing routines implemented at NOMAD \cite{mcdon;aca17} were used to generate diffraction patterns suitable for Rietveld refinements and real-space pair distribution function (PDF) data with a maximum momentum transfer of $Q_{\mathrm{max}}=30$~\AA$^{-1}$. Rietveld refinements were completed using TOPAS \cite{18_coelho} and fits to the PDF data were performed in PDFgui~\cite{farro;jpcm07}. These measurements allowed us to probe both the local and global structure of these materials. 

Neutron spectroscopy data were obtained from the direct-geometry time-of-flight instrument SEQUOIA \cite{Granroth_2010} at the SNS using polycrystalline samples of Ho$_2$GaSbO$_7$ and Ho$_2$ScSbO$_7$ with masses of 2.5~g and 4.5~g respectively. The samples were loaded in cylindrical Al cans with 1 atm of He gas and then mounted in a closed cycle refrigerator for these measurements. All SEQUOIA data were collected at 5~K with an incident energy of 150 meV or 4~meV using the fine Fermi chopper. The $T_0$ frequency, Fermi chopper frequency, and energy resolution at the elastic line (full-width half-maximum) were 90(30)~Hz, 600(120)~Hz, and 4(0.07)~meV for the $E_i =$~150(4)~meV dataset. These instrument settings allowed us to get a broad overview of the crystal-field spectrum and probe the splitting of the crystal-field ground state doublets for the Ho$^{3+}$ ions of each system.  

Temperature and field-dependent DC magnetization measurements were performed on a 7~T Quantum Design (QD) Magnetic Property Measurement System (MPMS3) SQUID magnetometer in vibrating-sample (VSM) mode.  Polycrystalline Ho$_2$GaSbO$_7$ and Ho$_2$ScSbO$_7$ samples were placed in polypropylene capsules and subsequently mounted in brass holders. Measurements were made using both the field-cooled and zero-field-cooled methodology. The temperature and field-dependent AC magnetization measurements were performed on a QD 9~T Dynacool Physical Property Measurement System (PPMS) employing the AC susceptibility option for the dilution fridge (ACDR). Each powder was cold pressed with a Carver press, and a portion of the resulting pellet with approximate dimensions of 1 $\times$ 1 $\times$ 0.5~mm$^3$ was adhered to a sapphire sample mounting post with a thin layer of Apiezon N-grease. All AC susceptibility measurements were collected under field-cooled (FC) conditions. 

Additional NPD data were collected on the high-resolution powder diffractometer HB-2A~\cite{10_garlea, 18_calder} at the High Flux Isotope Reactor (HFIR) of ORNL to investigate the magnetic diffuse scattering of polycrystalline Ho$_2$GaSbO$_7$ and Ho$_2$ScSbO$_7$. Polycrystalline Ho$_2$Ti$_2$O$_7$ was also measured to facilitate a direct comparison to the magnetic signal of a known DSI. Between 2.5-3~g of each sample was loaded into an Al can with 10 atm of He gas, which was subsequently mounted in a bottom-loading He-3 closed cycle refrigerator. The data were collected with a neutron wavelength of 2.41~\AA~and a collimation of open-21$'$-12$'$. Rietveld refinements were performed using the \texttt{FULLPROF} software suite~\cite{93_rodriguez} to confirm the known crystal structures of the materials and the magnetic diffuse scattering was analyzed with the reverse Monte Carlo method using the software \texttt{SPINVERT} \cite{12_paddison, 13_paddison}. 

The $T$-dependence of the low-energy spectra were measured on the cold triple axis spectrometer CTAX at the HFIR using the same \HGSO\ and \HSSO\ polycrystalline samples from HB-2A, which were mounted in the $^{3}$He insert of a liquid helium cryostat. The collimation settings were guide-open-80'-open. Constant-$Q$ scans were performed at several temperatures between 0.3 and 10 K with $Q =$~0.6~\AA$^{-1}$ and a final energy $E_f =$~4.8~meV to achieve an energy resolution of $\approx$0.27~meV (FWHM) at the elastic line. 

\section{Results and Discussion}

\subsection{Crystal Structures}

\begin{figure*}
\includegraphics[width=180mm]{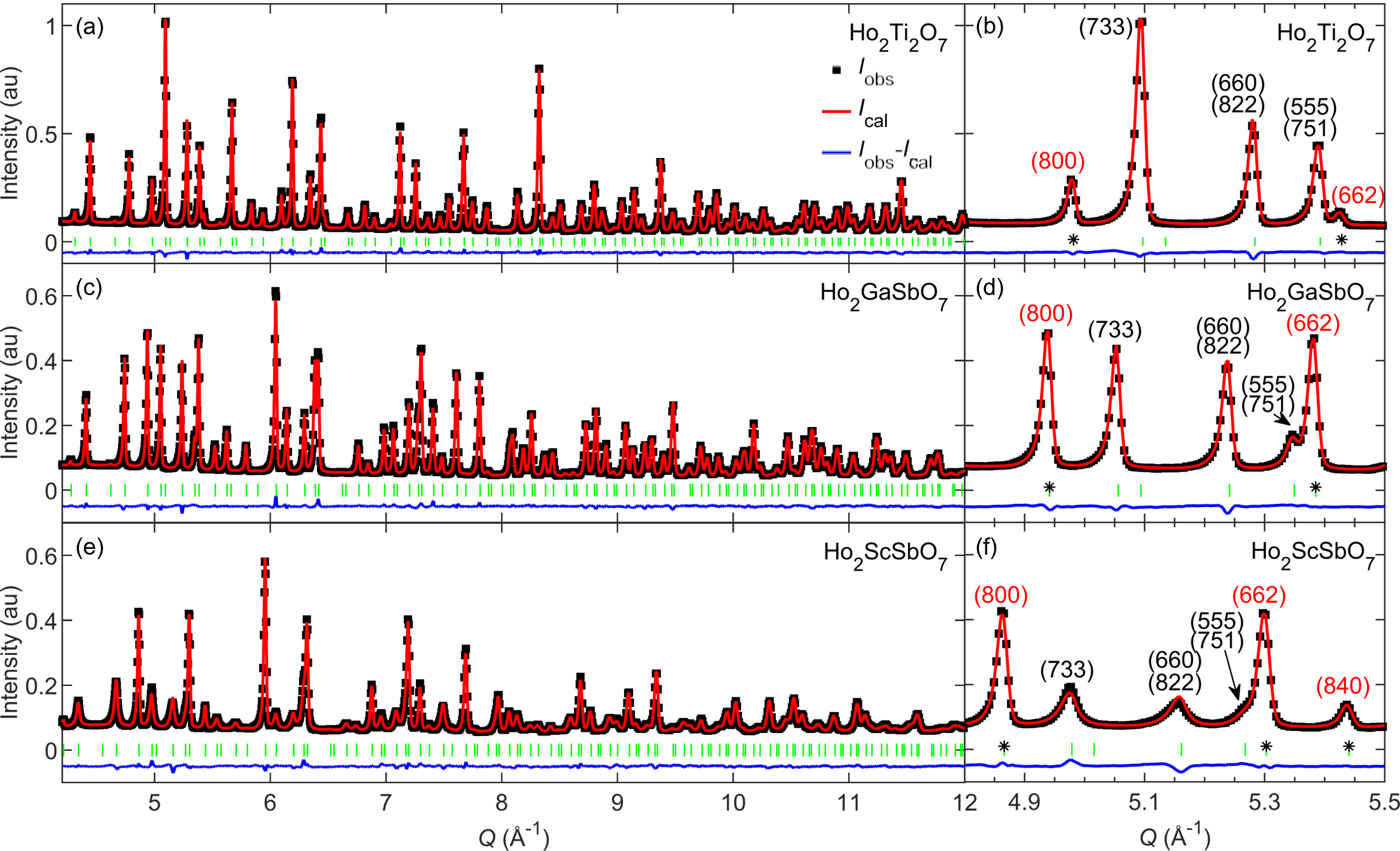}
 \caption{\label{fig:Rietveld} (a) Neutron powder diffraction pattern at $T =$~115~K for Ho$_2$Ti$_2$O$_7$ with the best Rietveld refinements superimposed on the data. The black squares represent the experimental data and the red curves represent the best Rietveld refinements. The difference plots are shown below the data and the expected Bragg peaks are indicated by tick marks. (b) The same data and Rietveld refinements shown over a limited $Q$-range to highlight selected nuclear Bragg peaks. The defect fluorite peaks are denoted by asterisks. (c-d) Similar plots for Ho$_2$GaSbO$_7$. (e-f) Similar plots for Ho$_2$ScSbO$_7$. While all the peaks are resolution-limited for Ho$_2$Ti$_2$O$_7$ and \HGSO, significant broadening is observed for the pyrochlore superlattice peaks in \HSSO.}
\end{figure*}

The rare-earth pyrochlore oxide structure $R_2B_2$O$_7$ is derived from the cubic fluorite structure $AB$O$_4$ by removing one eighth of the oxygen ions to create an array of ordered vacancies. For an intermediate cation radius ratio 1.46~$\le r_{R^{3+}}/r_{B^{4+}} \le$~1.82 \cite{ortiz2022traversing}, cation ordering across two different crystallographic sites is achieved to generate the $2\times2\times2$ superlattice pyrochlore structure. Although the cation radius rule is generally assumed to hold for mixed $B$-site systems, this has not been assessed in an exhaustive manner. For this reason, we begin our investigation of Ho$_2$GaSbO$_7$ and Ho$_2$ScSbO$_7$ by revisiting their crystal structures. 

The NPD patterns collected at 115~K using NOMAD's high-resolution backscattering detector bank (bank number 5) are presented in Fig.~\ref{fig:Rietveld} for both \HGSO\ and \HSSO. NPD data was also collected for Ho$_2$Ti$_2$O$_7$ to facilitate a straightforward comparison to a Ho pyrochlore with no $B$-site mixing. The best Rietveld refinements are superimposed on the data for all three systems, with the lattice parameters, fractional coordinates, and anisotropic displacement parameters provided in Table I. While the positions and intensities of the Bragg peaks were captured well using a simple pyrochlore model with instrument resolution-limited peak broadening for Ho$_2$Ti$_2$O$_7$ and \HGSO, there were significant deviations between the data and this model for \HSSO. Careful inspection of the NPD pattern revealed that the \HSSO\ pyrochlore superlattice peaks were significantly broader than the parent defect fluorite peaks, while all the peaks were resolution-limited for Ho$_2$Ti$_2$O$_7$ and \HGSO. This difference is highlighted in Figs.~\ref{fig:Rietveld}(b)(d)(f), where the indexed parent defect fluorite peaks are denoted by asterisks. To improve the \HSSO\ refinement, additional peak broadening beyond the contribution from the instrumental resolution function was incorporated into the model for the pyrochlore superlattice peaks only. This selective, intrinsic peak broadening implies that the Ho and Sc/Sb cation ordering in \HSSO\ has a finite correlation length $\zeta$. To estimate this value, we first fit the (733) pyrochlore superlattice peak and the (800) defect fluorite peak to Gaussian functions and then used the expression
\begin{equation}\label{HalfLorentzian}
\zeta = \frac{2\pi}{\sqrt{F^2_\mathrm{PS}-F^2_\mathrm{DF}}},
\end{equation}
where $F_\mathrm{PS}$ and $F_\mathrm{DF}$ are the full-width half-maximums of the (733) and (800) nuclear Bragg peaks, respectively. Note that this equation assumes both the instrumental resolution and the intrinsic peak broadening are Gaussian functions. A correlation length of 300~\AA~is obtained by this method, indicating that the structure is best described as pyrochlore up to approximately this length scale, but that the structural coherence of the ordered oxygen vacancies giving rise to the pyrochlore supercell is lost between regions of the crystal separated by more than this distance.

\begin{table*}[tb]
\centering
\caption{Structural parameters for Ho$_2$Ti$_2$O$_7$, Ho$_2$GaSbO$_7$, and Ho$_2$ScSbO$_7$ at $T=$~115~K extracted from the NOMAD neutron powder diffraction data.}

\subfloat[\HTO~at $T=115$ K, $R_{wp}=2.008$, $\mathrm{GOF}=2.851$]{
\begin{tabular}{lccccccccc}
\hline\hline
\multicolumn{5}{l}{Lattice parameter (\AA): 10.09035(5)} & \multicolumn{5}{r}{Space group: \textit{Fd$\bar{3}$m}} \\
\hline
Atom (Wyckoff) & $x$ & $y$ & $z$ & $u_{11}$ & $u_{22}$ & $u_{33}$ & $u_{12}$ & $u_{13}$ & $u_{23}$ \\
\hline
Ti (16c) & 0 & 0 & 0 &  0.00411(12) &0.00411(12) &0.00411(12) &-0.00039(13) &-0.00039(13) &-0.00039(13) \\
Ho (16d) & 0.5 & 0.5 & 0.5 & 0.00321(7)& 0.00321(7)& 0.00321(7)& -0.00119(5)& -0.00119(5)& -0.00119(5) \\
O1 (48f)  & 0.329543(18) & 0.125 & 0.125 & 0.00527(11) &0.00428(8) &0.00428(8) &0 &0 &0.00079(9) \\
O2 (8b)   & 0.375 & 0.375 & 0.375 & 0.00357(9)& 0.00357(9)& 0.00357(9) & 0 & 0 & 0  \\
\hline\hline
\end{tabular}
}\label{HTO}

\vspace{0.5em} 

\subfloat[\HGSO~at $T=115$ K, $R_{wp}=2.104$, $\mathrm{GOF}=2.651$]{
\begin{tabular}{lccccccccc}
\hline\hline
\multicolumn{5}{l}{Lattice parameter (\AA): 10.17178(5)} & \multicolumn{5}{r}{Space group: \textit{Fd$\bar{3}$m}} \\
\hline
Atom (Wyckoff) & $x$ & $y$ & $z$ & $u_{11}$ & $u_{22}$ & $u_{33}$ & $u_{12}$ & $u_{13}$ & $u_{23}$ \\
\hline
Ga/Sb (16c) &0&0&0 & 0.00355(5) &0.00355(5) &0.00355(5) &-0.00010(6) &-0.00010(6) &-0.00010(6) \\
Ho (16d) & 0.5 & 0.5 & 0.5 & 0.00787(10)& 0.00787(10) &0.00787(10) &-0.00331(8) &-0.00331(8) &-0.00331(8) \\
O1 (48f)  & 0.331369(19) & 0.125 & 0.125 & 0.00569(11) &0.00517(9) &0.00517(9) &0 &0 &0.00165(10) \\
O2 (8b)   & 0.375 & 0.375 & 0.375 & 0.00361(11) &0.00361(11) &0.00361(11) &0 &0 &0 \\
\hline\hline
\end{tabular}
}\label{HGSO}

\vspace{0.5em} 

\subfloat[\HSSO~at $T=115$ K, $R_{wp}=2.549$, $\mathrm{GOF}=2.897$]{
\begin{tabular}{lccccccccc}
\hline\hline
\multicolumn{5}{l}{Lattice parameter (\AA): 10.32970(13)} & \multicolumn{5}{r}{Space group: \textit{Fd$\bar{3}$m}} \\
\hline
Atom (Wyckoff) & $x$ & $y$ & $z$ & $u_{11}$ & $u_{22}$ & $u_{33}$ & $u_{12}$ & $u_{13}$ & $u_{23}$ \\
\hline
Sc/Sb (16c) & 0 & 0 & 0 & 0.00689(17) &0.00689(17) &0.00689(17) &-0.00075(17) &-0.00075(17) &-0.00075(17)
\\
Ho (16d) & 0.5 & 0.5 & 0.5 & 0.00693(19) &0.00693(19) &0.00693(19) &-0.00290(18) &-0.00290(18) &-0.00290(18)
\\
O1 (48f)  & 0.33466(7) & 0.125 & 0.125 & 0.0142(3) &0.0082(2) &0.0082(2) &0 &0 &0.0037(2)
 \\
O2 (8b)   & 0.375 & 0.375 & 0.375 & 0.0050(3) &0.0050(3) &0.0050(3) &0 &0 &0
 \\
\hline\hline
\end{tabular}
}\label{HSSO}
\end{table*}












\begin{figure}
	\includegraphics[width=86mm]{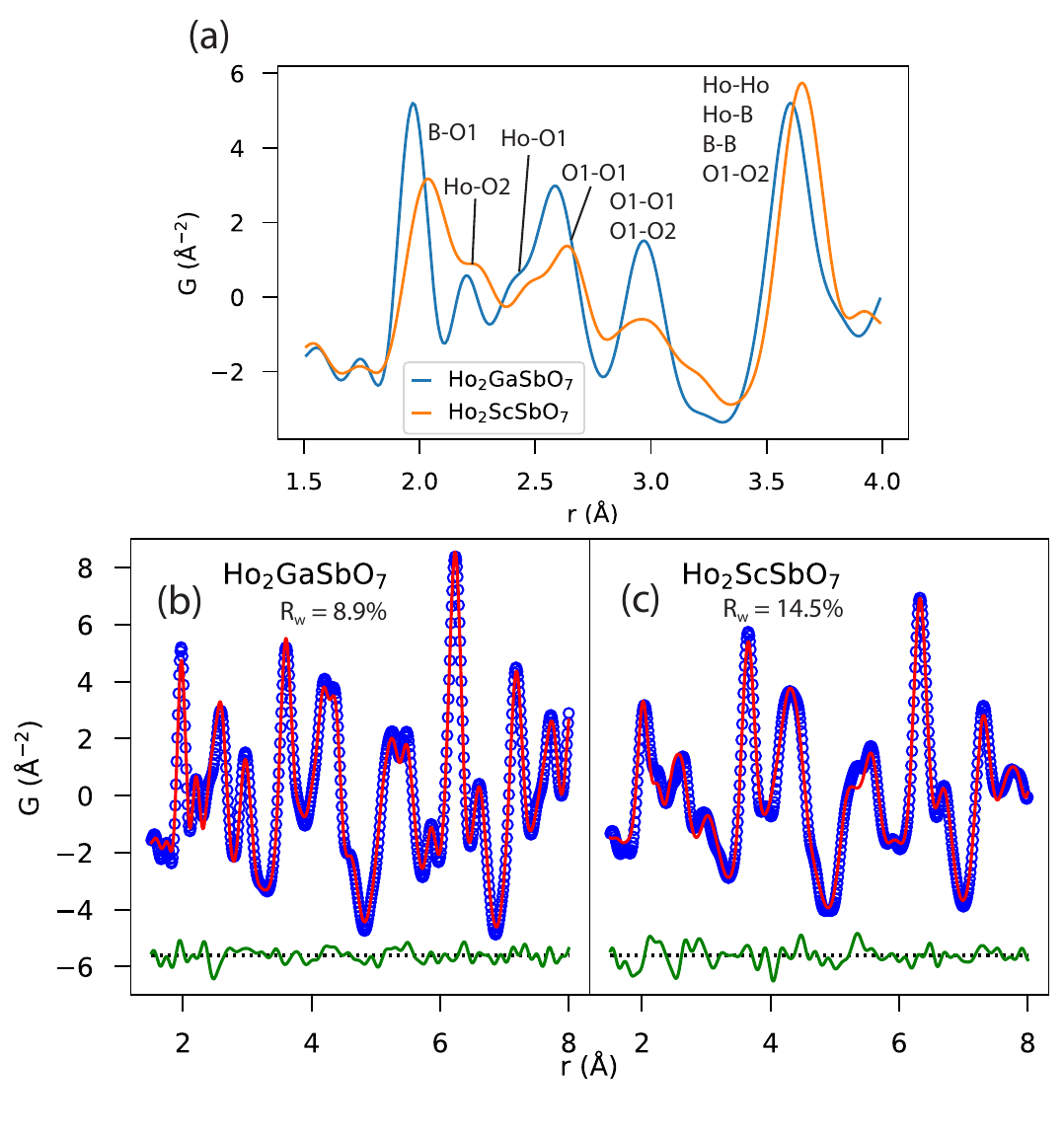}
	\caption{\label{fig:PDF}(a) Experimental PDF patterns for \HGSO\ and \HSSO\ highlighting differences in the local structure. Atom pairs contributing to specific peaks are indicated. Broader peaks in \HSSO\ are a consequence of increased local disorder due to the larger ionic radius size difference. (b) PDF fit for \HGSO, with open blue symbols showing the experimental data, the red curve the best fit, and the green curve the fit residual, offset vertically for clarity. The $R_\mathrm{w}$ value displayed is a goodness of fit metric, with smaller values corresponding to a better fit. (c) Same as (b), but for \HSSO.}
\end{figure}

For \HGSO, $r_{\mathrm{Ho}^{3+}}/r_{B_\mathrm{av}}=$~1.66, where $B_\mathrm{av}$ is the average $B$-site radius, places the system well in the pyrochlore stability field, so the success of the Rietveld refinement is not surprising. There is also a negligible difference between the $B$-site ionic radii (0.62~\AA\ for Ga$^{3+}$ vs 0.60~\AA\ for Sb$^{5+}$). While $r_{\mathrm{Ho}^{3+}}/r_{B_\mathrm{av}}=$~1.51 for \HSSO\ should also stabilize a pyrochlore structure with long-range cation ordering in principle, the large ionic radius difference of Sc$^{3+}$ (0.745~\AA) and Sb$^{5+}$ generates significant lattice strain that can promote structural distortions and hence disrupt the long-range cation ordering typical of pyrochlores. A similar effect has been observed previously in stuffed Ho$_{2+x}$Ti$_{2-x}$O$_{7-x/2}$ \cite{07_lau}, where the excess Ho ions are introduced on the $B$-site. For increasing $x$, the large size mismatch between Ho$^{3+}$ (0.90~\AA) and Ti$^{4+}$ (0.605~\AA) first destabilizes the pristine pyrochlore structure via phase separation into regions with long and short-range cation ordering before evolving into a phase with short-range cation ordering only for Ho$_2$TiO$_5$ ($x = 0.67$). 

To examine the disorder of these mixed $B$-site pyrochlores at the local level, we turn to PDF analysis. The pair distribution function $G(r)$ is plotted as a function of distance $r$(\AA) for the two samples at 115~K in Fig.~\ref{fig:PDF}. As seen in panel (a), the PDF peaks corresponding to many of the near-neighbor atom pairs are much broader for \HSSO\ than for \HGSO. This demonstrates that the large $B$-site ionic radius size difference in \HSSO\ results in a broad distribution of bond lengths, as one might expect. On the one hand, even the peaks that include oxygen but not a $B$-site cation, e.g. Ho-O and O-O peaks, are significantly broader in \HSSO\ than in \HGSO, indicating that the oxygen atoms surrounding the $B$-sites are displaced significantly from their average position due to the different sizes of the $B$-site cations. On the other hand, the peak around 3.6~\AA\ that is dominated by Ho-Ho, $B$-$B$, and Ho-$B$ pairs is comparably sharp for both compounds, which means that the $B$-site size difference is accommodated primarily by the oxygen positions, with much less disorder in the Ho- and $B$-sublattice positions.

We further characterize the local disorder in each compound by fitting a model of the average pyrochlore structure to the PDF data, shown in Fig.~\ref{fig:PDF}(b-c). In this model, all $B$ sites are occupied by a virtual atom with the average scattering length of the two distinct $B$-site cations consistent with the average cubic symmetry of the pyrochlore structure. The model works well for \HGSO, as seen by the relatively flat difference curve shown in green, but it performs much worse for \HSSO, with large deviations between the data and model seen up to about 6~\AA. This is also reflected in the $R_\mathrm{w}$ goodness-of-fit value, which is much larger (indicating a worse fit) for \HSSO. Beyond 6~\AA, the average-structure model works well for both compounds, indicating that distortions of the local structure are most significant below this distance. Finally, we note that for \HSSO, the atomic displacement parameters (ADPs) for the O1 atoms are much larger ($\approx$ 0.012~\AA$^{2}$) than those for the O2 atoms ($\approx$ 0.005~\AA$^{2}$). The ADPs set the width of the peaks corresponding to specific pairs of atoms, reflecting the effect of thermal vibrations and/or non-thermal broadening due to intrinsic disorder. The much larger value for the O1 atoms indicates that they experience larger local displacements than do the O2 atoms. This is expected given that the O2 atoms coordinate only the Ho site, whereas O1 atoms also coordinate the $B$ site with mixed ionic species. These results suggest that the size mismatch of the $B$-site ions is accommodated almost entirely by local distortions to the O1 positions. In contrast, the ADPs for the O1 and O2 atoms in \HGSO\ are small and comparable to each other (approximately 0.005~\AA$^2$ and 0.004~\AA$^2$, respectively), confirming that unusually large random displacements are absent from both oxygen sites when the ionic radius size mismatch is small.

The PDF data are also consistent with the NPD analysis indicating a finite correlation length for the pyrochlore superstructure in \HSSO. When fitting the pyrochlore and defect fluorite models to the \HSSO\ PDF data over relatively short ranges, such as the first 20~\AA, the pyrochlore model provides a far better fit than the defect fluorite model (9.3\% compared to 34.4\%). This confirms that the overall local structure is pyrochlore, albeit with significant local displacements of the O1 atoms as discussed previously. However, when fitting to data ranges at much higher $r$ values, such as 80 -- 100~\AA, the pyrochlore and defect fluorite models yield fits of comparable quality (40.2\% compared to 42.4\%; we note that the high $R_\mathrm{w}$ values are due to artificial distortions known to exist at high $r$ in PDF data collected on NOMAD~\cite{olds;aca18}). Boxcar fits with a sliding data range of width 20~\AA\ ranging from [0, 20]~\AA\ to [80, 100]~\AA\ show that the models become comparable for distances beyond about 80~\AA, though this is likely an underestimate of the crossover length scale due to the high-$r$ artifacts in the data. Nevertheless, the overall trend agrees qualitatively with the previous assessment that the pyrochlore superstructure has a finite correlation length on the order of tens of nanometers.

\subsection{Single Ion Properties}

Ho$_2$Ti$_2$O$_7$ \cite{00_rosenkranz, 16_ruminy} and other Ho-based pyrochlores \cite{00_matsuhira, 02_kadowaki, 12_hallas} are known to have strong Ising anisotropy with the moments constrained to local $\langle111\rangle$ directions. A common approach to determine the crystal-field parameters and the $g$-tensor of the ground state doublet for these materials is to fit the energy levels and integrated intensities of the crystal-field excitations measured with neutron spectroscopy using the single ion Hamiltonian appropriate for pyrochlores,
\begin{equation}\label{HamiltonianHo}
\mathcal{H}_\mathrm{si} = B^0_2\hat{O}^0_2 + B^0_4\hat{O}^0_4 + B^3_4\hat{O}^3_4 + B^0_6\hat{O}^0_6 + B^3_6\hat{O}^3_6 + B^6_6\hat{O}^6_6,
\end{equation}
where $B_l^m$ and $\hat{O}_l^m$ represent the crystal-field parameters and the corresponding operators in Stevens notation. 

\begin{figure}
	\includegraphics[width=86mm]{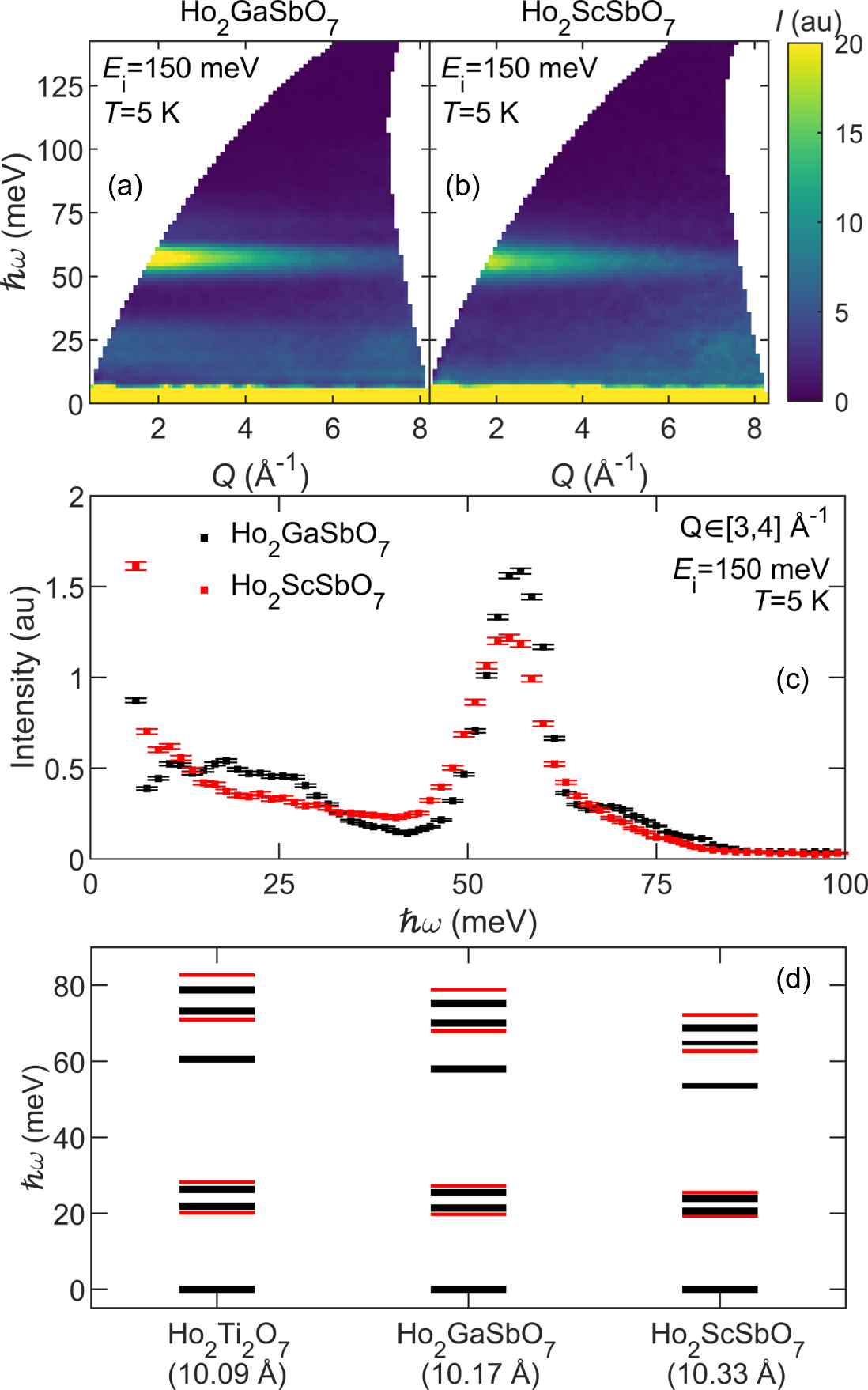}
	\caption{\label{fig:CEF} (a-b) Color contour plots of the scattering intensity $I(Q,E)$ for Ho$_2$GaSbO$_7$ and Ho$_2$ScSbO$_7$ from SEQUOIA with $E_i =$~150~meV and $T =$~5~K. The spectrum of both systems is dominated by a broad band of magnetic excitations centered just below 60~meV, which corresponds well to the most intense crystal-field excitation measured previously for Ho$_2$Ti$_2$O$_7$ \cite{00_rosenkranz, 16_ruminy}. (c) A constant-$Q$ cut of the scattering intensity with an integration range of $Q =$~[3,4]~\AA$^{-1}$ for both systems. Broad crystal-field excitations are observed in both cases, with peaks shifted to lower energies for Ho$_2$ScSbO$_7$ due to the larger lattice constant. The crystal-field excitations are generally broader for Ho$_2$ScSbO$_7$ due to its enhanced structural disorder. (d) The measured crystal-field energy levels for Ho$_2$Ti$_2$O$_7$ and the calculated crystal-field energy levels for Ho$_2$GaSbO$_7$ and Ho$_2$ScSbO$_7$ using a scaling analysis. The crystal-field doublets and singlets are indicated by black and red lines respectively.}
 \end{figure} 

\begin{table}[b]
    \centering
    \caption{Determination of the Stevens parameters (in units of meV) for Ho$_{2}$GaSbO$_{7}$ and Ho$_{2}$ScSbO$_{7}$ from scaling arguments using the crystal-field parameters reported for Ho$_{2}$Ti$_{2}$O$_{7}$.}
    \label{table:scaling}
    \begin{tabular}{ c|c|c|c } 
        \hline
        $B_l^m$  & Ho$_{2}$Ti$_{2}$O$_{7}$ & Ho$_{2}$GaSbO$_{7}$ & Ho$_{2}$ScSbO$_{7}$ \\ \hline \hline
        $B^0_2$ & $-0.0781$     & $-0.0762$ & $-0.0728$ \\ 
        $B^0_4$ & $-0.00117$   & $-0.00112$ & $-0.00104$ \\ 
        $B^3_4$ & $-0.00803$   & $-0.00771$ & $-0.00714$ \\ 
        $B^0_6$ & $-0.00000707$ & $-0.00000668$ & $-0.00000600$ \\ 
        $B^3_6$ & $0.000103$    & $0.0000974$ & $0.0000874$ \\ 
        $B^6_6$ & $-0.000133$   & $-0.000125$ & $-0.000113$ \\ 
        \hline
    \end{tabular}
\end{table}

Neutron spectroscopy data collected on SEQUOIA at 5~K, with an incident energy $E_i =$~150~meV, are presented in Fig.~\ref{fig:CEF}(a) and (b) for \HGSO\ and \HSSO\ respectively. The scattering intensity is plotted as a function of momentum transfer $Q$ and energy transfer $\hbar\omega$ using a color contour map. The dominant feature in both spectra is a broad crystal-field excitation centered around 60~meV that is also visible in the constant-$Q$ cuts shown in Fig.~\ref{fig:CEF}(c). This mode is reminiscent of previous measurements on Ho$_2$Ti$_2$O$_7$, where the most intense crystal-field level was found at 61~meV \cite{16_ruminy}. Additional spectral weight is visible both below and above the $\sim$60~meV mode for both mixed $B$-site systems, which is not surprising since the remaining crystal-field levels associated with the Ho$^{3+}$ $J = 8$ ground state multiplet in Ho$_2$Ti$_2$O$_7$ are found between 20 and 30~meV and 70 and 85~meV. Although these levels should be doublets in principle due to the high point symmetry of the Ho ions, evidence for a vibronic bound state has now been observed in Ho$_{2}$Ti$_{2}$O$_{7}$ that splits the 61~meV level by $\approx$~3~meV \cite{18_gaudet}. Similarly to previous neutron spectroscopy work on other mixed $B$-site pryochlores such as Nd$_2$GaSbO$_7$ \cite{21_gomez}, the significant broadening of the crystal-field levels beyond expectations for the vibronic bound state effect precludes extracting the (average) Ho$^{3+}$ crystal-field parameters via conventional analysis. 

\begin{figure*}
	\includegraphics[width=175mm]{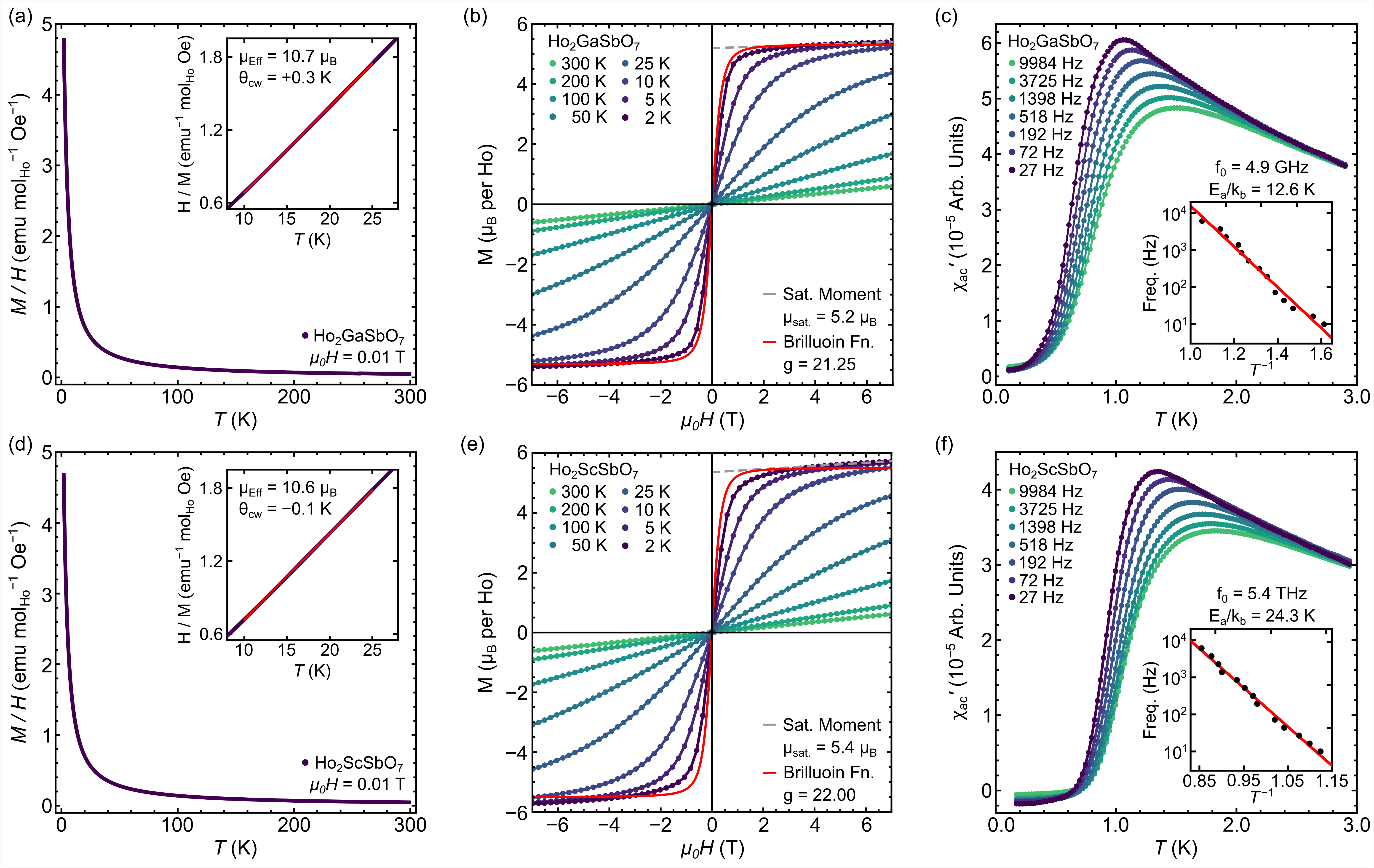}
	\caption{\label{fig:BulkCharact} (a) Magnetic susceptibility, plotted as $M/H$, vs temperature in an applied magnetic field of 0.01~T for Ho$_2$GaSbO$_7$. Curie-Weiss behavior is observed down to the base temperature of 2 K. The Curie-Weiss fitting results are shown in the inset. (b) $M$ vs $H$ at selected temperatures for Ho$_2$GaSbO$_7$. The saturation magnetization value is close to expectations for Ising pyrochlores. (c) The real part of the AC susceptibility $\chi_{ac}'$ vs temperature for Ho$_2$GaSbO$_7$, which reveals complete spin freezing. Results of the Arrhenius analysis are shown in the inset. (d-f) Similar data and results for Ho$_2$ScSbO$_7$.}
 \end{figure*}

A scaling analysis procedure to estimate the average crystal-field parameters can be used instead. This approach has been employed for many other pyrochlore systems (see e.g. \cite{Bertin_2012}) when neutron spectroscopy data is not available or the number of observables required for the conventional analysis is not sufficient. The scaled crystal-field parameters for the two mixed $B$-site pyrochlore systems can be calculated as follows: 
\begin{equation}\label{HamiltonianHo}
B_l^m(\mathrm{Ho}_2B_2\mathrm{O}_7) = \frac{a^{l+1}(\mathrm{Ho}_2\mathrm{Ti}_2\mathrm{O}_7)}{a^{l+1}(\mathrm{Ho}_2B_2\mathrm{O}_7)}B_l^m(\mathrm{Ho}_2\mathrm{Ti}_2\mathrm{O}_7)
\end{equation}
where $a$ denotes the lattice constant for each compound. Using the lattice constants for the three pyrochlores provided in Table~I and the crystal-field parameters for Ho$_2$Ti$_2$O$_7$ reported in Ref.~\cite{16_ruminy}, we obtain the crystal-field parameters for \HGSO\ and \HSSO\ listed in Table~II. The crystal-field energy schemes for the three systems extracted from this analysis are illustrated in Fig.~\ref{fig:CEF}(d). As the lattice constant of the system increases, the energy levels shift downwards. Nonetheless, all three systems have large energy gaps to the first excited crystal-field level (20.10, 19.75, and 19.13~meV for Ho$_2$Ti$_2$O$_7$, \HGSO, and \HSSO\ respectively) and therefore an effective spin-$\frac{1}{2}$ model should be valid for all of them below 10~K. The magnetic anisotropy in this regime will then be governed by the ground state doublet $g$-tensor, which is given by $g_\parallel =$~19.56 (19.61) for \HGSO\ (\HSSO). The scaling analysis indicates that the two mixed $B$-site systems maintain strong Ising anisotropy with $J_z \approx \pm 8$ ground state doublet wavefunctions. We note that this simple scaling analysis neglects the expected splitting of the crystal-field ground state doublet due to the $B$-site disorder, which we discuss in Section~III D.

\subsection{Magnetic Ground States}\label{sub:mag_gs}

\begin{figure*}
\includegraphics[width=175mm]{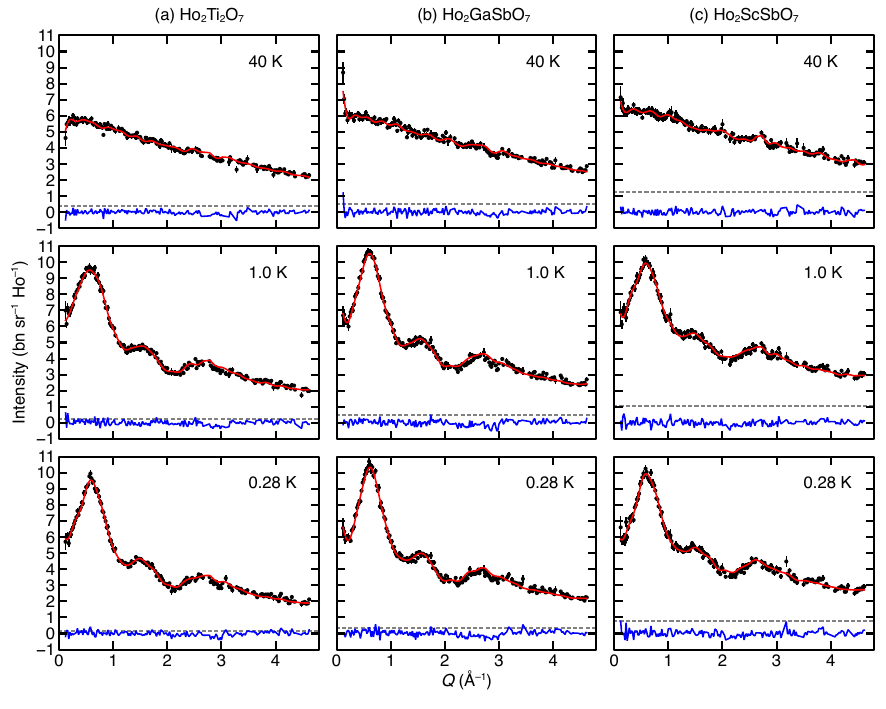}
 \caption{\label{fig:rmc} Magnetic diffuse scattering data (black circles), reverse Monte Carlo fits (red curves), and data--fit (blue curves) for (a) Ho$_2$Ti$_2$O$_7$, (b) Ho$_2$GaSbO$_7$, and (c) Ho$_2$ScSbO$_7$. For each compound, three temperatures are shown (0.28, 1.0, and 40\,K, as labeled in the figure). Grey dashed lines in each panel indicate the refined constant background level for each data set.
 }
\end{figure*}

The effect of the significant structural differences for \HGSO\ and \HSSO\ on their magnetic ground states is now examined. The dc magnetic susceptibility $\chi = M/H$ of both samples measured from 2-300~K in an applied field of 0.01~T is presented in Fig.~\ref{fig:BulkCharact}(a,d). No signs of magnetic ordering are observed in the data. The insets depict the same data in the low-$T$ regime only with fits to the Curie-Weiss law superimposed. The effective moment and Curie-Weiss temperature for \HGSO\ (\HSSO) are $10.7$($10.6$)~$\mu_\mathrm{B}$ and $0.3$($-0.1$)~K respectively. The large effective moments comparable to Ho$^{3+}$ isolated ion values are consistent with the crystal field scaling analysis discussed above and the small Curie-Weiss temperatures are in line with previous results for other Ho pyrochlores. The field-dependence of the magnetization at selected temperatures is shown in Fig.~\ref{fig:BulkCharact}(b,e) with the best-fit Brillouin function superimposed on each of the 2~K datasets. The saturation magnetization at 2~K for both systems is close to half of the 10~$\mu_B$  value for a $J_z = \pm 8$ ground state for Ho$^{3+}$, as expected for powder samples of Ising pyrochlores \cite{00_bramwell}. The slightly larger saturated magnetization values likely arise from small deviations away from the $J_z = \pm 8$ ground state for the Ho$^{3+}$ ions due to the continuous distribution of low-lying crystal-field levels associated with the $B$-site disorder. 

AC susceptibility measurements with driving frequencies spanning multiple decades were conducted on both samples. The real part of the AC susceptibility $\chi^{\prime}_\mathrm{ac}$ at selected frequencies over a $T$-range of 0.02~K to 3~K is depicted in Fig.~\ref{fig:BulkCharact}(c,f). Both samples are characterized by a frequency-dependent peak followed by a sharp drop to zero with decreasing temperature. The spin freezing transition temperature, defined by the peak position in the imaginary part of the AC susceptibility $\chi^{\prime\prime}_\mathrm{ac}$, is 0.62~K (0.90~K) for \HGSO\ (\HSSO) using the 10~Hz data. With increasing frequency, the peak positions in both the real and imaginary components of the AC susceptibility shift to higher temperatures. This behavior is a known hallmark of glassy transitions. The frequency-dependence of the imaginary-component peak position can be fitted to an Arrhenius law,
\begin{equation}\label{Arrhenius}
f = f_0 \mathrm{exp}(-E_a/ k_\mathrm{B} T),
\end{equation}
which yields $f_0 =$~4.9~GHz (5.4~THz) and an activation energy $E_a/k_\mathrm{B} =$~12.6~K (24.3~K). The parameters from the Arrhenius analysis are broadly consistent with the results for other Ho-based pyrochlores with dipolar spin ice ground states \cite{00_matsuhira, 12_hallas}. For these systems, $E_a$ is related to the energy cost of the single spin flip required to generate two emergent magnetic monopole excitations. 

The strong Ising anisotropy inferred from our crystal-field scaling analysis and saturation magnetization values, along with the complete spin freezing signatures identified in the bulk characterization data, provide strong support for dipolar spin ice ground states in \HGSO\ and \HSSO. Another important signature of this exotic state is a magnetic diffuse scattering pattern with a characteristic oscillatory $Q$-dependence \cite{02_kadowaki, 04_mirebeau, 12_hallas}. To look for this behavior in our materials, we carried out low-temperature neutron powder diffraction on the HB-2A constant-wavelength diffractometer between 0.28~K and 40~K in \HGSO, \HSSO, and Ho$_2$Ti$_2$O$_7$ for reference. We first verified that all three systems retained the pyrochlore structure over this low-temperature range via Rietveld refinements of their diffraction patterns.

Our main diffuse scattering results for all three systems are shown in Fig.~\ref{fig:rmc}. To estimate this scattering, we processed the data as follows. Data were corrected for absorption \cite{79_hewat}, and were placed in absolute intensity units by normalizing to the nuclear Bragg profile \cite{25_paddison}. The calculated incoherent scattering level was subtracted and nuclear Bragg peaks were removed by excluding data points where the calculated Bragg intensity exceeded a threshold value. As anticipated, the magnetic diffuse scattering obtained in this way for Ho$_2$Ti$_2$O$_7$ resembles previous measurements \cite{04_mirebeau, 12_hallas}. At the highest temperature of 40~K, the magnetic scattering decreases with increasing $Q$, as expected for an uncorrelated paramagnet due to the $Q$-dependence of the magnetic form factor. As the temperature is lowered, spin-ice correlations develop and generate the scattering pattern observed at 1~K, with a well-defined peak below $Q = 1$~\AA$^{-1}$ as expected for the dipolar spin-ice model \cite{12_paddison}. The peaks and valleys in the pattern become slightly more distinct as the temperature is lowered to 0.28~K. The magnetic diffuse scattering patterns for \HGSO\ and \HSSO\ are nearly identical to the Ho$_2$Ti$_2$O$_7$ pattern, although there is small variation in the amplitudes of the peak and valley features.  

The reverse Monte Carlo (RMC) technique was used to model the magnetic diffuse scattering data in order to test quantitatively the hypothesis of  spin-ice ground states in all three systems. The RMC method has now been used successfully to analyze powder-averaged magnetic diffuse scattering data of many frustrated magnets \cite{12_hallas} and software tools such as \texttt{SPINVERT} \cite{13_paddison} are available. The RMC method does not require a magnetic Hamiltonian, but rather refines a large configuration of spin orientations to obtain the best fit to experimental data. In our RMC analysis of the magnetic diffuse scattering for the three Ho-based pyrochlores, our main assumptions are that spins occupy an undistorted pyrochlore lattice, spins behave as classical Ising variables that are constrained to point along the local $\langle111\rangle$ pyrochlore axes, and all spins have equal length. The scattering patterns were calculated from the RMC spin configurations according to the general expression provided elsewhere \cite{64_blech, 13_paddison}. Refinements were performed using supercells of $5\times 5 \times 5$ crystallographic unit cells (2000~spins) with periodic boundary conditions, and were run for 200 proposed flips per spin, after which no further improvement in fit quality was observed. At each iteration of the refinement, an overall intensity scale factor and a constant background term were optimized to match the data. The latter term accounts for any residual background that was not subtracted in the data correction, as well as for multiple scattering and for incoherent scattering that can arise from very small amounts of water in the sample or apparatus.

As illustrated in Fig.~\ref{fig:rmc}, the RMC fits for the three samples at all temperatures show excellent agreement with the experimental data. The fraction of Ho tetrahedra that obey the ice rules can be calculated from these fits. At the base-temperature of 0.28~K, we find values of 93\%, 93\%, and 85\% for Ho$_2$Ti$_2$O$_7$, \HGSO, and \HSSO\ respectively. The slightly-reduced value for \HSSO\ may be related to the local structural distortions and/or the short-range cation ordering discussed above. Notably, these estimates represent lower bounds because RMC fitting generally produces the most disordered spin configurations compatible with experimental data \cite{07_tucker}. 

\subsection{Low-Energy Magnetic Excitations}

\begin{figure*}
\includegraphics[width=175mm]{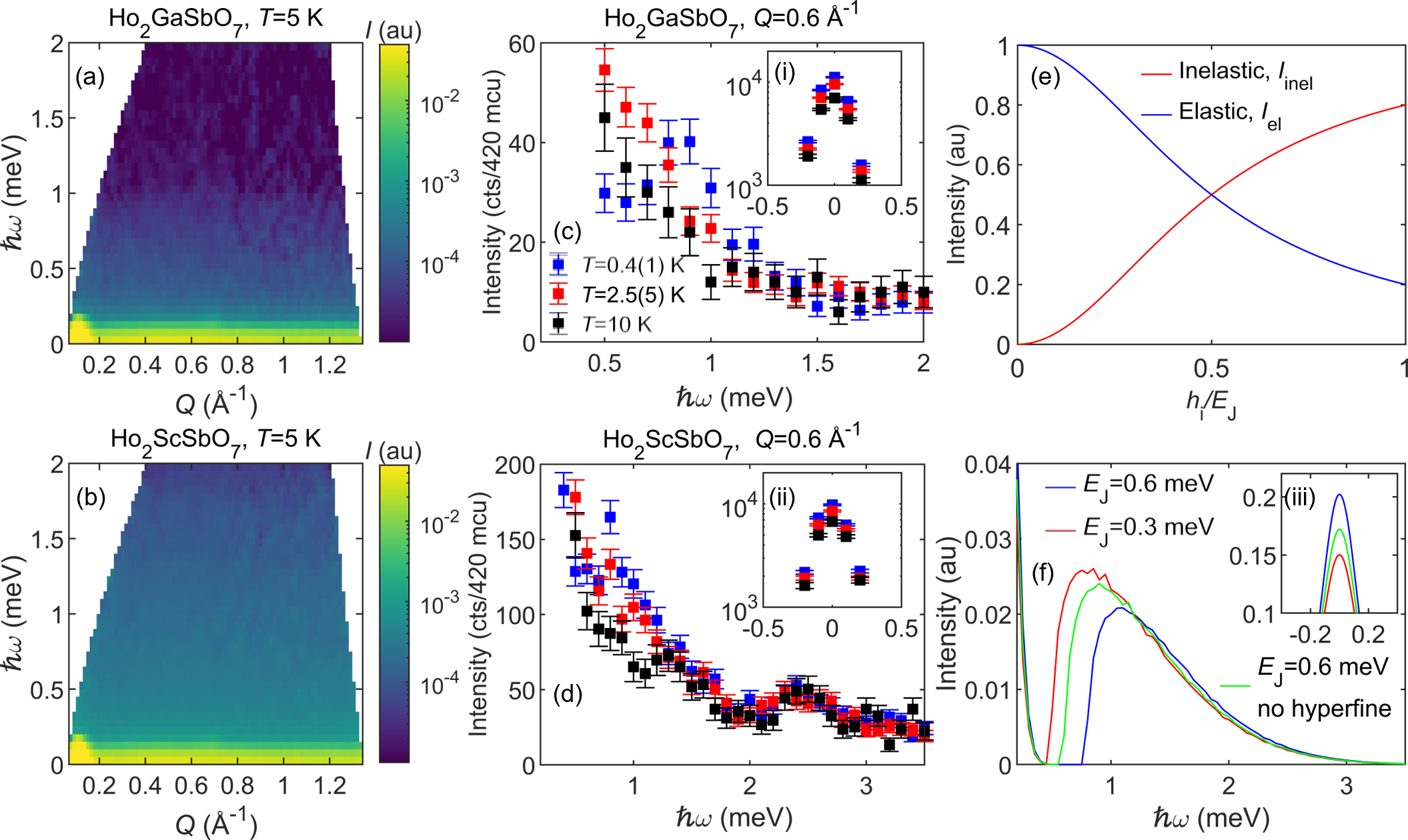}
\caption{\label{fig:quasielastic} (a,b) Color contour intensity maps of inelastic neutron scattering from polycrystalline \HGSO\ and \HSSO, respectively, plotted as functions of momentum transfer $Q$ and energy transfer $\hbar\omega$, collected on SEQUOIA with $E_i=4$ meV at $T=5$~K. (c,d) Constant-$Q = 0.6$~\AA$^{-1}$ scans measured at $T = 0.4$ K (averaged from measurements at $T = 0.3$~K and $0.5$~K), 2.5~K (averaged from $T = 2$~K, 2.5~K, and 3~K), and 10~K on CTAX for \HGSO\ (2.93 g) and \HSSO\ (2.66 g), respectively, collected on CTAX. Note that 1 monitor count unit (mcu) is approximately 1 s for elastic scattering. Data in panel (b,d) are rescaled by a factor of $\sim1.06$ to account for the mass difference between the two samples, ensuring the intensities correspond to the same number of Ho$^{3+}$ ions. Insets (i,ii) show the corresponding elastic lines. (e) Evolution of the calculated neutron scattering cross-sections for the elastic channel ($I_{\mathrm{el}}$) and inelastic channel ($I_{\mathrm{inel}}$), centered at energy transfer $\Delta_{\mathrm{gap}}$, as a function of the ratio $h_i/E_J$ between the uniform transverse field and the mean-field spin-flip energy, using Eq.~\ref{Iel_inel}. (f) Simulated energy-dependent magnetic scattering intensities based on the mean-field model described in Eqs.~\ref{Eqn:Htot}–\ref{Eqn:CEF}, calculated at $T=0$. A half-Gaussian distribution of transverse fields $h_i$ 
is assumed [Eq.~\ref{hgdist}] with a width parameter $\sigma = 0.5$~meV. Blue and red curves correspond to two representative values of $E_J = 0.6$~meV and 0.3~meV, respectively. The green curve shows the simulated spectrum for $E_J = 0.6$~meV in the absence of the hyperfine interaction [Eq.~\ref{Eqn:Hhf_ss}]. The elastic contribution has been convolved with the instrumental energy resolution (FWHM = 0.27~meV) for visualization. The peak of the simulated elastic intensity is shown in inset~(iii).}
 \end{figure*}

With the dipolar spin ice ground states for these materials now established, we present their low-energy excitation spectra. Neutron spectroscopy experiments on the two mixed $B$-site pyrochlore systems reveal clear experimental signatures consistent with the random transverse field model introduced in Eq.~\ref{Eq:Ham3} and discussed in Refs.~\cite{17_savary, 18_benton, 19_pardini}. Time-of-flight inelastic neutron scattering measurements performed on SEQUOIA with an incident energy of $E_i = 4$~meV provide an overview of the low-energy spectra of both systems at $T = 5$~K, as shown in the color contour intensity maps plotted as functions of momentum transfer $Q$ and energy transfer $\hbar\omega$ in Fig.~\ref{fig:quasielastic}(a,b). The data reveal inelastic magnetic scattering extending up to $\approx$~2~meV for both compounds. 

Constant-$Q$ scans with $Q = 0.6$~\AA$^{-1}$, collected at $T = 0.4$, 2.5, and 10~K on CTAX, are shown in Fig.~\ref{fig:quasielastic}(c,d), highlighting the temperature dependence of the inelastic magnetic response. At $T = 10$~K, the spectra exhibit quasi-elastic scattering extending up to $\approx$~2~meV for \HGSO\ and $\approx$~3.5~meV for \HSSO. At higher energy transfers, the inelastic signal approaches the typical CTAX background level of 1-2 counts per minute. The \HSSO~data additionally exhibit a weak peak centered just below 2.5~meV, which shows little $T$-dependence down to 0.4~K. Its origin cannot be determined conclusively: a phonon or crystal-field excitation cannot be confirmed due to the lack of $Q$-dependent measurements on CTAX and the presence of a 3~meV spurious feature in the SEQUOIA data that extends down to $\approx$~2.5~meV. Upon cooling below the spin-ice regime around $T \approx 1$~K, a well-defined peak centered at $\hbar\omega = 0.9(1)$~meV emerges in both compounds.

The inelastic magnetic spectra observed in both \HGSO\ and \HSSO\ are absent in pristine dipolar spin ice systems such as \HTO ~\cite{09_clancy}. These excitations arise from crystal-field transitions between split singlets of the Ho$^{3+}$ non-Kramers ground-state doublet, induced by $B$-site disorder that breaks the local $D_{3d}$ symmetry. Following previous studies on Pr-based pyrochlores~\cite{16_petit, Martin2017}, we introduce a mean-field single-site approximation to the random transverse-field model (Eq.~\ref{Eq:Ham3}) to describe the temperature dependence of the inelastic spectra. In this framework, the effect of local disorder on the non-Kramers doublet is represented by a site-dependent transverse field $h_i$ acting on each Ho$^{3+}$ ion. 

The system is thus described, within a mean-field approximation, by a two-level Hamiltonian at each site $i$,
\begin{equation}
\mathcal{H}^{\mathrm{mf}}_{0,i} =
\begin{pmatrix}
0 & -h_i \\
-h_i & E_J
\end{pmatrix},
\label{HMF}
\end{equation}
written in the basis of pseudospin states denoted $\ket{+}$ and $\ket{-}$ and aligned along the local $\hat{z}\in \langle 111\rangle$ axis. Here, $E_J$ represents the energy cost to flip a pseudospin from $\hat{z}$ to $-\hat{z}$, which, in the absence of disorder, corresponds to the creation of a magnetic monopole pair. Within this mean-field model, $E_J$ depends on the configuration of neighboring Ho$^{3+}$ pseudospins. For a given site $i$ with nearest neighbors $j$, the spin-flip energy is given by $E_J = 2\sum_{j} J_{\parallel} \langle \sigma_j^z \rangle$. In the case where the surrounding pseudospins adopt a `2-in-2-out' configuration, one finds $E_J = 4J_{\parallel}$. Diagonalizing this Hamiltonian yields a split doublet consisting of a ground state $\ket{\mathrm{GS}}$ and excited state $\ket{\mathrm{EX}}$ with energy gap $\Delta_{\mathrm{gap}}$,
\begin{align}
\begin{split}
\ket{\mathrm{GS}} &= a_1 \ket{-} + b_1 \ket{+}, \\
\ket{\mathrm{EX}} &= a_2 \ket{-} + b_2 \ket{+}, \\
\Delta_{\mathrm{gap}} &= \sqrt{E_J^2 + 4h_i^2},
\end{split}
\end{align}
where $a_1$, $b_1$, $a_2$, and $b_2$ are normalized coefficients, satisfying $|a_1|^2 + |b_1|^2 = 1$ and $|a_2|^2 + |b_2|^2 = 1$. 

In systems with non-Kramers ground-state doublets~\cite{Chen2017DiracMonopoles}, the local $\hat{z}$ spin component transforms as a dipole and hence couples to the neutron spin, while the other spin components transform as quadrupoles and do not couple to the neutron spin. The projected spin operator $J_z$ within the non-Kramers doublet is proportional to the pseudospin operator $\sigma_z$~\cite{16_petit}, so the neutron scattering intensities in the elastic and inelastic channels can be expressed as
\begin{align}
\begin{split}
I_{\mathrm{el}} &\propto \left| \bra{\mathrm{GS}} \sigma^z \ket{\mathrm{GS}} \right|^2 = (b_1^2 - a_1^2)^2=\frac{E_J^2}{\Delta_{\mathrm{gap}}^2}, \\
I_{\mathrm{inel}}(\Delta_{\mathrm{gap}}) &\propto \left| \bra{\mathrm{EX}} \sigma^z \ket{\mathrm{GS}} \right|^2 = (b_1 b_2 - a_1 a_2)^2=\frac{4h_i^2}{\Delta_{\mathrm{gap}}^2}. \label{Iel_inel}
\end{split}
\end{align}

In the classical spin ice limit ($h_i = 0$), one finds $a_2 = b_1 = 1$, indicating that the ground state is fully aligned along the local $\hat{z}$ direction and satisfies the  ice rule on each tetrahedron. In this case, $I_{\mathrm{inel}} = 0$ and all magnetic spectral weight is confined to the elastic line. In the opposite limit ($h_i \gg J_\parallel$), the pseudospin lies nearly perpendicular to the local easy axis, resulting in a singlet ground state $\ket{\mathrm{GS}}\rightarrow1/\sqrt{2}(\ket{+}+\ket{-})$ with vanishing elastic scattering ($I_{\mathrm{el}} \rightarrow 0$) and complete transfer of spectral weight to the inelastic channel at energy transfer $\hbar \omega = \Delta_{\mathrm{gap}}$. The evolution of $I_{\mathrm{el}}$ and $I_{\mathrm{inel}}$ for a fixed $E_J$ and varying $h_i$ is shown in Fig.~\ref{fig:quasielastic}(e). 

To model the effect of structural disorder in real Ho pyrochlore samples, two further ingredients are needed. First, it is necessary to assume a specific distribution of local transverse fields. We consider a half-Gaussian form,
\begin{align}
p(h) = \frac{2}{\sqrt{2\pi\sigma^2}} \exp\!\left(-\frac{h^2}{2\sigma^2}\right), \qquad h \geq 0,
\label{hgdist}
\end{align}
with $p(h) = 0$ for $h < 0$ and a width parameter $\sigma = 0.5$~meV. Second, the nuclear hyperfine interaction \cite{ramirez1994nuclear,Dun2020_spinFragmentation} 
was neglected in Eq.~\ref{Eq:Ham3}, but can be important in Ho-based systems with quasi-doublet crystal-field ground states \cite{Dun2020_spinFragmentation}. The nuclear hyperfine interaction introduces an additive term in the single-ion Hamiltonian \cite{ramirez1994nuclear,Dun2020_spinFragmentation} given by
\begin{equation}
\mathcal{H}^{\mathrm{mf}}_{\mathrm{hf},i}=A\boldsymbol{J}\cdot\boldsymbol{I} = 8A \sigma^z \otimes I_z,\label{Eqn:Hhf_ss}
\end{equation}
where $\sigma^z$ is the Pauli matrix; $I_z$ is an $8\times8$ diagonal matrix with eigenvalues $m_I = \pm7/2, \pm5/2, \pm3/2, \pm1/2$, corresponding to the $I = 7/2$ nuclear spin manifold of $^{165}$Ho. Here $A = 3.36~\mu\mathrm{eV} \approx 0.039$~K~\cite{ramirez1994nuclear} denotes the bare hyperfine coupling constant for Ho. At the mean-field level, the hyperfine interaction influences the spectroscopic response in two main ways. First, it broadens both the elastic ($I_{\mathrm{el}}$) and inelastic ($I_{\mathrm{inel}}$) components of the scattering intensity through the splitting of the nuclear spin degeneracy. Second, it acts cooperatively with the $E_J$ term, enhancing the excitation gap $\Delta_{\mathrm{gap}}$ and favoring more spectral weight in the elastic channel.

The total mean-field Hamiltonian can therefore be expressed as
\begin{equation}
\mathcal{H}^{\mathrm{mf}}_i = \mathcal{H}^{\mathrm{mf}}_{0,i} \otimes \hat{I}_{8\times8} + \mathcal{H}^{\mathrm{mf}}_{\mathrm{hf},i},\label{Eqn:Htot}
\end{equation}
where $\hat{I}_{8\times8}$ denotes the $8\times8$ identity matrix. The inelastic spectra associated with crystal-field transitions can then be simulated using
\begin{equation}
\frac{d^2\sigma}{d\Omega\,d\omega} \propto 
\sum_{m,n} p_m\, 
\big|\langle n |\, \sigma^z \otimes \hat{I}_{8\times8} \,| m \rangle \big|^2 
\delta\!\big(\hbar\omega - (E_{n} - E_{m})\big),\label{Eqn:CEF}
\end{equation}
where the summation runs over all transitions between the eigenstates $\ket{m}$ and $\ket{n}$, and $p_m = \exp(-E_{m}/k_B T)/Z$ represents the Boltzmann population factor of the initial state $\ket{m}$. Representative simulation results at zero temperature are shown in Fig.~\ref{fig:quasielastic}(f) for two characteristic values of $E_J$. The choice $E_J \approx 0.6$~meV corresponds to a nearest-neighbor exchange strength of $J_{\parallel} = 1.75$~K for Ho$_2$Ti$_2$O$_7$~\cite{Zhou2012} and a fully developed `2-in–2-out' configuration. For comparison, the simulated spectrum for $E_J \approx 0.6$~meV in the absence of the hyperfine interaction is also shown.

As discussed previously for the constant $h_i$ case, increasing $E_J$ produces an inelastic spectrum with a larger gap $\Delta_{\mathrm{gap}}$ and reduced spectral weight, accompanied by a corresponding transfer of spectral weight to the elastic channel. The simulated response exhibits a sharp onset at energy $\sim E_J$, followed by a broad tail extending to higher energies. This trend is expected, as the inelastic matrix element $I_{\mathrm{inel}}$ increases with the ratio $h_i / E_J$. Importantly, the high-energy tail is relatively insensitive to variations in $E_J$, as it arises from sites where the transverse field dominates, i.e., $h_i \gg E_J$. In practice, the energy dependence of $I_{\mathrm{inel}}$ is further shaped by the detailed form of the transverse field distribution $p(h)$ and the dispersion of the gapped excitations induced by $J_{\parallel}$ beyond the scope of the mean-field approximation.

Within this framework, the 0.9~meV peak observed in both compounds below approximately 1~K can be interpreted as a consequence of crystal-field level restructuring as the system enters the spin-ice regime. As the `2-in–2-out' spin configurations emerge at low temperatures, the nearest-neighbor exchange field strengthens, leading to an increase in spin-flip energy, 
$E_J = 2\sum_{j} J_{\parallel} \langle \sigma_j^z \rangle \rightarrow 4J_{\parallel}$. This enhanced exchange field promotes alignment of the pseudospins along the local $\hat{z}$ direction. Consequently, the spectral weight associated with sites where $h_i \sim J_{\parallel}$ is transferred from inelastic crystal-field transitions to the elastic line, accompanied by an upward shift in the lower bound of the crystal-field excitation energies to $E_J \approx 0.6$~meV. This behavior is qualitatively consistent with our experimental observations. It is important to note, however, that the observed increase in elastic intensity at lower temperatures (Fig.~\ref{fig:quasielastic} (i,ii)) cannot be straightforwardly interpreted in terms of the predicted temperature dependence of the crystal-field intensity. This is because our measurements were performed at a fixed momentum transfer $Q$, while the $Q$-dependence of the magnetic scattering evolves as spin-ice correlations gradually develop. We also observe stronger inelastic scattering in \HSSO\ compared to \HGSO, indicating a larger average transverse field in \HSSO. This observation is reasonable given that the doped $B$-site ion Ga has an ionic radius more similar to that of the original Sb ion, whereas Sc has a significantly larger radius. The larger mismatch in ionic size introduced by Sc leads to greater local distortion from the ideal $D_{3d}$ symmetry, thereby generating a stronger average transverse field.

\section{Conclusions}
Introducing controlled disorder on the non-magnetic $B$-site of Ho-based pyrochlores provides a powerful means of testing theoretical predictions of the random transverse-field Ising model, which can stabilize a disorder-induced quantum spin-ice state over a broad parameter range. In this work, we synthesized and characterized the mixed $B$-site compounds \HGSO\ and \HSSO\ using bulk measurements and multiple neutron-scattering techniques. These two systems offer a natural contrast: the $B$-site ionic radii in \HGSO\ are nearly identical, whereas \HSSO\ exhibits a substantial mismatch between Sc$^{3+}$ and Sb$^{5+}$. This difference significantly impacts their structural properties. Although both materials crystallize in the pyrochlore structure, the superlattice Bragg peaks in \HSSO\ are not resolution-limited, indicating a reduced correlation length for cation ordering.

Bulk thermodynamic measurements and magnetic diffuse scattering confirm that both materials retain the defining signatures of a dipolar spin-ice ground state, including low-temperature spin freezing. However, we also observe clear evidence of enhanced local structural distortions in \HSSO\ and show how they impact the low-energy magnetic dynamics. The larger distortions generate a broader distribution of crystal-field singlet–singlet energy splittings, which can be effectively modeled as stronger random transverse fields acting on each Ho site. These structural variations manifest directly in the magnetic excitation spectra: whereas neutron scattering from the parent compound Ho$_2$Ti$_2$O$_7$ remains within the elastic resolution in the spin-ice regime, both \HGSO\ and \HSSO\ display substantial inelastic magnetic scattering, with \HSSO\ exhibiting a larger excitation bandwidth consistent with its stronger disorder.

Taken together, our results demonstrate that dipolar spin-ice physics in Ho pyrochlores is remarkably robust to non-magnetic $B$-site disorder, while also showing that such disorder provides a natural route to tuning disorder-induced quantum fluctuations in non-Kramers frustrated magnets.

\section{acknowledgments}
We acknowledge useful discussions with O. Benton. B.R.O. (synthesis, bulk properties measurements and analysis) gratefully acknowledges support from the U.S. Department of Energy (DOE), Office of Science, Basic Energy Sciences, Materials Sciences and Engineering Division. B.A.F. (pair distribution function analysis) was supported by the U.S. National Science Foundation, Division of Materials Research, LEAPS-MPS program through award No. 2418438. A portion of this research used resources at the Spallation Neutron Source and High Flux Isotope Reactor, which are DOE Office of Science User Facilities operated by Oak Ridge National Laboratory. The beam time was allocated to CG-4C (CTAX) on proposal number IPTS-30989.1 and IPTS-31600.1 and BL-17 (SEQUOIA) on proposal number IPTS-28453.1 and IPTS-29949.1. Additional beam time was allocated to BL-1B (NOMAD) on proposal number IPTS-28963.1 and HB-2A (POWDER) on IPTS-31023.1. S.D.W., G.P. and P.M.S. acknowledge support from the U.S. Department of Energy (DOE), Office of Basic Energy Sciences, Division of Materials Sciences and Engineering under Grant No. DE-SC0017752. A portion of this work used facilities supported via the UC Santa Barbara NSF Quantum Foundry funded via the Q-AMASE-i program under award DMR-1906325.

\section{Data Availability}
The data that support the findings of this article are openly available \cite{zenodo}.

\bibliography{main}
\bibliographystyle{apsrev4-2}

\end{document}